\newcommand{\resection}[1]{\setcounter{equation}{0}\section{#1}}

\documentclass[a4paper,11pt]{article}
\usepackage{amsfonts}

\usepackage{amsmath}

\newcommand{\EQ}{\begin{equation}}
\newcommand{\EN}{\end{equation}}
\newcommand{\bea}{\begin{eqnarray}}
\newcommand{\eea}{\end{eqnarray}}

\begin{document}

\setcounter{page}{0} \topmargin0pt \oddsidemargin5mm \renewcommand{%
\thefootnote}{\arabic{footnote}}\newpage \setcounter{page}{0}
\begin{titlepage}
\begin{flushright}
SISSA 93/2007/EP
\end{flushright}
\vspace{0.5cm}
\begin{center}
{\large {\bf Isomorphism of critical and off-critical operator spaces}\\
{\bf in two-dimensional quantum field theory}}\\

\vspace{1.8cm}
{\large Gesualdo Delfino$^{a,b}$ and Giuliano Niccoli$^{c\,\,*}$} \\
\vspace{0.5cm}
{\em ${}^a$International School for Advanced Studies (SISSA)}\\
{\em via Beirut 2-4, 34014 Trieste, Italy}\\
{\em ${}^b$INFN sezione di Trieste}\\
{\em ${}^c$LPTM, Universit\'e de Cergy-Pontoise }\\
{\em 2 avenue Adolphe Chauvin, 95302 Cergy-Pontoise, France}\\
{\em E-mail: delfino@sissa.it, giuliano.niccoli@desy.de}\\
\end{center}
\vspace{1.2cm}

\renewcommand{\thefootnote}{\arabic{footnote}}
\setcounter{footnote}{0}

\begin{abstract}
\noindent
For the simplest quantum field theory originating from a non-trivial fixed
point of the renormalization group, the Lee-Yang model, we show
that the operator space determined by the particle dynamics
in the massive phase and that prescribed by conformal symmetry at criticality
coincide.
\end{abstract}

\vspace{8cm}
$^*$\,Present address: DESY Theory, Notkestr. 85, 22607 Hamburg, Germany.
\end{titlepage}

\newpage

\resection{Introduction}
The local operators of a quantum field theory form an infinite-dimensional
linear space which the transformation properties under space-time and internal
symmetries naturally decompose into subspaces, each containing infinitely many
operators with assigned spin and charge. For the generic quantum field theory
describing a renormalization group trajectory flowing out of a non-trivial
fixed point, a further, quantitative characterization of the operator space is
a very difficult task. Particularly relevant is the characterization of the
operators by their scaling dimension, as determined by the short distance
behavior of the two point functions. The spectrum of the scaling dimensions,
as well as the number of operators sharing the same dimension, are
distinguishing data of the theory which remain, however, out of reach in the
generic case.

Two-dimensional quantum field theory enjoys in this respect a special status.
Here, the infinite-dimensional nature of conformal symmetry allows the
solution of fixed point theories \cite{BPZ}, revealing in particular a
decomposition of the operator space into families containing operators with
integer-spaced scaling dimensions. The number of families, the scaling
dimensions and the degeneracy within each integer level are all known. Two
non-negative integers, $l$ and $\bar{l}$, determine, through their difference
and their sum, the spin and (up to a constant characteristic of the family)
the scaling dimension of an operator.

Perturbative considerations \cite{Taniguchi,Alyosha,GM} suggest that, up to
symmetry breaking effects, such a structure should survive a perturbation of
the fixed point leading to a massive theory. The existence in two dimensions
of massive integrable theories provides the chance of testing
non-perturbatively this conjecture, but also poses a problem which is
intriguing and challenging at the same time. Indeed, integrable massive
theories are solved {\em on-shell}, through the determination of the exact
$S$-matrix, and any information about the operators needs to be extracted from
the particle dynamics. The monodromy properties and the singularity structure
provide a set of equations \cite{KW,Smirnov} for the matrix elements of the
operators on asymptotic states (form factors) which have the $S$-matrix as
their only input, and whose space of solutions is expected to coincide with
the operator space of the theory.

The latter issue was first addressed by Cardy and Mussardo in \cite{CM}, where
the simplest case, that of the thermal Ising model, was investigated.  They
showed that the number of solutions of the form factor
equations with spin $s$ and with the mildest asymptotic behavior at high
energy coincides with the number of chiral operators (i.e. having $\bar{l}=0$)
fixed by conformal field theory for spin $s=l$ at the ultraviolet fixed
point\footnote{Of course a similar analysis can be performed for the operators
with $l=0$ (antichiral).}. Mildest asymptotic behavior is
a reasonable conjecture for the chiral operators, which are the most relevant
among the operators with given spin.

While clearly supporting the idea of the isomorphism of
critical and off-critical operator spaces, this original study deferred to
subsequent investigations some important issues. First of all, the counting
of generic, non-chiral, operators requires the introduction in the form
factor approach of some information about the levels, which are no longer
uniquely specified by the spin. Moreover, as already pointed out in \cite{CM},
the case of the thermal Ising model is ``deceptively simple'' due to its
equivalence with a free fermionic theory\footnote{The spin sector
considered in \cite{CM}, however, is non-trivial due to non-locality with
respect to the fermions.}. In particular, all the form factor solutions can be
generated by repeated action of the conserved quantities, a circumstance which
does not persist for the generic integrable theory.

The problem of considering interacting integrable theories was
tackled by Koubek \cite{Koubek}, who extended the analysis of Cardy and
Mussardo for the {\it conjectured} chiral solutions to several massive
deformations
of minimal conformal models. She also performed a more general counting of the
form factor solutions according to the asymptotic behavior for given spin,
obtaining very suggestive formal relations with the characters which in
conformal field theory specify the structure of the operator families. The
inability to establish a connection with the levels $l$ and $\bar{l}$,
however, prevented her from showing the isomorphism between the conformal and
massive operator spaces. It was shown in \cite{Smirnovcounting,BBS,JMT} how
this type of counting can be extended to the sine-Gordon model and its
restrictions.

In this paper we show explicitly for the massive Lee-Yang model that the space
of solutions of the form factor equations decomposes into subspaces labeled by
a pair of non-negative integers $(l,\bar{l})$ related to the spin and to the
asymptotic
behavior of the form factors at high energy. We then show that the dimension
of each subspace exactly coincides with the dimension of the subspace of the
conformal operator space with levels $(l,\bar{l})$, proving in this way the
isomorphism between the critical and off-critical operator spaces. The choice
of the Lee-Yang model for a first time proof is obvious: on one hand, this
model is fully representative of generic integrable quantum field theories
with respect to the features which are of interest here (it is a massive,
interacting theory originating from a non-trivial fixed point of the
renormalization group, with an operator space which cannot be entirely
generated by repeated
action of conserved quantities on lowest level operators); on the other, it
minimizes the technicalities and best illustrates the essential points due to
the presence of the minimal number of operator families (two) and of a single
species of massive particles.

The paper is organized as follows. After recalling in the next section the
structure of the operator space of the Lee-Yang model at criticality, we turn
in section~3 to the analysis of the space of solutions of the form factor
equations in the massive theory. The comparison between the critical and
off-critical operator spaces is then performed in section~4. Section~5 contains
few final remarks while some technical parts of the proof are detailed in two
appendices.

\resection{The conformal operator space}

At a critical point the operators undergo the general conformal field theory
classification \cite{BPZ}. A scaling operator $\Phi (x)$ is first of all
labeled by a pair $(\Delta _{\Phi },\bar{\Delta}_{\Phi })$ of conformal
dimensions which determine the scaling dimension $X_{\Phi }$ and the spin
$s_{\Phi }$ as
\begin{eqnarray}
&&X_{\Phi }=\Delta _{\Phi }+\bar{\Delta}_{\Phi } \\
&&s_{\Phi }=\Delta _{\Phi }-\bar{\Delta}_{\Phi }.
\end{eqnarray}
There exist operator families associated to the lowest weight
representations of the Virasoro algebra
\begin{equation}
\lbrack L_{n},L_{m}]=(n-m)L_{n+m}+\frac{c}{12}n(n^{2}-1)\delta _{n+m,0}\,.
\label{virasoro}
\end{equation}
The $L_{n}$'s generate the conformal transformations associated to the
complex variable $z=x_{1}+ix_{2}$, with the central charge $c$ labeling the
conformal theory. The same algebra, with the same value of $c$, holds for
the generators $\bar{L}_{n}$ of the conformal transformations in the
variable $\bar{z}=x_{1}-ix_{2}$. The $L_{n}$'s commute with the $\bar{L}_{m}$%
's. Each operator family consists of a primary operator $\Phi _{0}$ (which
is annihilated by all the generators $L_{n}$ and $\bar{L}_{n}$ with $n>0$)
and infinitely many descendant operators obtained through the repeated
action on the primary of the Virasoro generators. A basis in the space of
descendants of $\Phi_0$ is given by the operators
\begin{equation}
L_{-i_{1}}\ldots L_{-i_{I}}\bar{L}_{-j_{1}}\ldots \bar{L}_{-j_{J}}\,\Phi _{0}
\label{descendants}
\end{equation}
with
\begin{eqnarray}
&&0<i_{1}\leq i_{2}\leq \ldots \leq i_{I} \\
&&0<j_{1}\leq j_{2}\leq \ldots \leq j_{J}\,.
\end{eqnarray}
The {\em levels}
\begin{equation}
(l,\bar{l})=\left( \sum_{n=1}^{I}i_{n}\,,\sum_{n=1}^{J}j_{n}\right)
\end{equation}
determine the conformal dimensions of the descendants (\ref{descendants}) in
the form
\begin{equation}
(\Delta ,\bar{\Delta})=(\Delta _{\Phi _{0}}+l,\bar{\Delta}_{\Phi _{0}}+\bar{l%
})\,.
\end{equation}

In general the number of independent operators at level $(l,\bar{l})$ is
$p(l)p(\bar{l})$, $p(l)$ being the number of partitions of $l$ into positive
integers. This number, however, is reduced in a model dependent way in presence
of degenerate representations
associated to primary operators $\phi _{r,s}$ which possess a vanishing
linear combination of descendant operators (null vector) when $l$ or $\bar{l}
$ equals $rs$. So the dimensionality of the subspace with levels
$(l,\bar{l})$ in the family of $\Phi_0$ is usually written as
$d_{\Phi_0}(l)d_{\Phi_0}(\bar{l})$, with the integers $d_{\Phi_0}(l)$ which can
be encoded in the rescaled {\em character}
\begin{equation}
\chi_{\Phi_0}(q)=\sum_{l=0}^{\infty }d_{\Phi_0}(l)q^{l}\,.
\label{character}
\end{equation}

The conformal field theory with the smallest operator content is the
minimal model $\mathcal{M}_{2,5}$, with central charge $c=-22/5$, possessing
only two primary operators: the identity $I=\phi_{1,1}=\phi_{1,4}$
with conformal dimensions $(0,0)$, and the operator $\varphi=\phi_{1,2}=
\phi_{1,3}$ with conformal dimensions $(-1/5,-1/5)$. The negative values of
the central charge and of $X_\varphi$ show that the model does not satisfy
reflection-positivity. It goes under the name of Lee-Yang model because it
describes (see \cite{Cardy}) the universal properties of the edge
singularity of the zeros of the partition function of the Ising model in an
imaginary magnetic field \cite{YL,LY,Fisher}.

The characters for the two operator families of the Lee-Yang model are
\cite{Feigen-Fuchs,R-C,Christe}
\begin{equation}
\chi _{I}(q)=\prod_{n=0}^{\infty }\frac{1}{\left( 1-q^{2+5n}\right) \left(
1-q^{3+5n}\right) }
\end{equation}
\begin{equation}
\chi _{\varphi }(q)=\prod_{n=0}^{\infty }\frac{1}{\left( 1-q^{1+5n}\right)
\left( 1-q^{4+5n}\right) }\,.
\end{equation}
They also enjoy the following ('fermionic' \cite{Fermionic-representations})
representation based on the Rogers-Ramanujan identities \cite{R-Rj-identities}:
\begin{equation}
\chi _{I}(q)=G_{-1}\text{, \ \ \ }\chi _{\varphi }(q)=G_{0},
\end{equation}
where
\begin{equation}
G_{p}=\sum_{k=0}^{\infty }\frac{q^{k(k-p)}}{(q)_{k}},  \label{G_M}
\end{equation}
\begin{equation}
(q)_{k}=\prod_{i=1}^{k}(1-q^{i})\,.
\end{equation}
This representation, when compared with the definition (\ref{character}),
yields the following expansions for the dimensions of the spaces of chiral
descendants of level $l$
\begin{equation}
d_{I}(l)=\sum_{N=0}^{\infty }P(N,l-N(N+1)),\text{ \ \ }d_{\varphi
}(l)=\sum_{N=0}^{\infty }P(N,l-N^{2}),
\label{d-i-phi}
\end{equation}
where $P(N,M)$ is the number of the partitions of the non-negative integer $M$
into the integers $1,2,\ldots,N$; it is generated by
\begin{equation}
\frac{1}{(q)_{N}}=\sum_{M=0}^{\infty }P(N,M)q^{M}.
\label{generating-partition}
\end{equation}
We set
\begin{equation}
P(0,0)=1,\text{ \ \ }P(N,M)=0\text{ \ \ for }N\geq 0\text{ and }M<0,\text{ \
\ }P(0,M)=0\text{ \ \ for }M>0;
\end{equation}
notice that $P(N,0)=1$.
The generating functions $G_{p}$ satisfy the recursion relation
\begin{equation}
G_{p}=G_{p-1}+q^{1-p}G_{p-2},
\end{equation}
which implies
\begin{equation}
\chi _{I}(q)+\chi _{\varphi }(q)=G_{1},  \label{char-I-phi}
\end{equation}
and leads to
\begin{equation}
d_{I}(l)+d_{\varphi }(l)=\sum_{N=0}^{\infty }P(N,l-N(N-1)).
\label{dim-tot(l,0)}
\end{equation}

The occurrence of different representations for the characters of rational
conformal field theories is a general phenomenon. Fermionic representations
have been derived for classes of rational conformal theories defined as cosets
of affine Lie algebras \cite{Fermionic-representations}
and in particular for the series of non-unitary minimal models $M(2,2p+3)$
\cite{Fermionic-Rep-M}. These representations follow by a
quasi-particle interpretation based on the Bethe Ansatz description and give
alternative representations of the characters with respect to the
Feigin-Fuchs-Felder construction \cite{Feigin-Fuchs-Felder construction}.
Their derivation uses generalizations
of the Rogers-Ramanujan identities (the Gordon-Andrews identities \cite
{G-A-identities}).

\resection{Operators in the massive theory}

A renormalization group trajectory originating from the Lee-Yang
conformal point is obtained perturbing the model ${\cal M}_{2,5}$ with
its only non-trivial primary operator $\varphi$. This gives the massive
Lee-Yang model with action
\begin{equation}
\mathcal{A}=\mathcal{A}_{CFT}+g\int d^2x\,\varphi(x)\,.  \label{action}
\end{equation}
It follows from the general results of \cite{Taniguchi} about perturbed
conformal field theories that the theory (\ref{action}) belongs to the class
of integrable quantum field theories. These are characterized by the existence
of an infinite number of conserved quantities which induces the
complete elasticity and factorization of the scattering processes, and
allows the exact determination of the $S$-matrix \cite{ZZ}. It was shown in
\cite{CMyanglee} that the massive Lee-Yang model has a mass spectrum containing
a single species of neutral particles $A(\theta)$ with two-body scattering
determined
by the amplitude\footnote{The on-shell two-momentum of a particle of mass $m$
is parameterized by a rapidity variable $\theta$ as $(p^{0},p^{1})=
(m\cosh \theta,m\sinh \theta)$. In (\ref{s}) $\theta$ denotes the rapidity
difference of the colliding particles.}
\begin{equation}
S(\theta )=\frac{\tanh \frac{1}{2}\left( \theta +\frac{2i\pi }{3}\right) }{%
\tanh \frac{1}{2}\left( \theta -\frac{2i\pi }{3}\right) }\,,  \label{s}
\end{equation}
which, due to factorization, specifies the full $S$-matrix.
The bound state pole located at $\theta =2i\pi /3$ corresponds to the fusion
process $AA\rightarrow A$ and has the residue
\begin{equation}
\mbox{Res}_{\theta =2i\pi /3}S(\theta )=i\Gamma ^{2},
\end{equation}
where $\Gamma =i2^{1/2}3^{1/4}$ is the three-particle coupling; the fact that
$\Gamma$ is purely imaginary is again related to the lack of
reflection-positivity \cite{CMyanglee}.

Within integrable quantum field theory the operators
are constructed determining their matrix elements on the asymptotic particle
states. All the matrix elements of a given local operator $\Phi (x)$ can be
obtained from the $n$-particle form factors
\begin{equation}
F_{n}^{\Phi }(\theta _{1},\ldots ,\theta _{n})=\langle 0|\Phi (0)|\theta
_{1}\ldots \theta _{n}\rangle ,  \label{form factors}
\end{equation}
where $|0\rangle$ denotes the vacuum (i.e. zero-particle) state.
The form factors satisfy a set of functional equations taking into account
the spin $s_\Phi$ of the operator, the monodromy properties under analytic
continuation in rapidity space and the pole singularities associated to bound
states and annihilation processes \cite{KW,Smirnov}. For the Lee-Yang model
these equations read
\begin{eqnarray}
&&F_{n}^{\Phi }(\theta _{1}+\alpha ,\ldots ,\theta _{n}+\alpha )=e^{s_{\Phi
}\alpha }F_{n}^{\Phi }(\theta _{1},\ldots ,\theta _{n})  \label{fn0} \\
&&F_{n}^{\Phi }(\theta _{1},\ldots ,\theta _{i},\theta _{i+1},\ldots ,\theta
_{n})=S(\theta _{i}-\theta _{i+1})\,F_{n}^{\Phi }(\theta _{1},\ldots ,\theta
_{i+1},\theta _{i},\ldots ,\theta _{n})  \label{fn1} \\
&&F_{n}^{\Phi }(\theta _{1}+2i\pi ,\theta _{2},\ldots ,\theta
_{n})=F_{n}^{\Phi }(\theta _{2},\ldots ,\theta _{n},\theta _{1})  \label{fn2}
\\
&&\mbox{Res}_{\theta ^{\prime }=\theta }\,F_{n+2}^{\Phi }(\theta ^{\prime }+%
\frac{i\pi }{3},\theta -\frac{i\pi }{3},\theta _{1},\ldots ,\theta
_{n})=i\Gamma \,F_{n+1}^{\Phi }(\theta ,\theta _{1},\ldots ,\theta _{n})
\label{fn4}\\
&&\mbox{Res}_{\theta ^{\prime }=\theta +i\pi }\,F_{n+2}^{\Phi }(\theta
^{\prime },\theta ,\theta _{1},\ldots ,\theta _{n})=i\left[
1-\prod_{j=1}^{n}S(\theta -\theta _{j})\right] F_{n}^{\Phi }(\theta
_{1},\ldots ,\theta _{n})\,,  \label{fn3}
\end{eqnarray}
with $S(\theta)$ and $\Gamma$ specified above. The space of solutions of these
equations is linear in the operators and is expected to coincide with the
infinite-dimensional operator space of the massive theory. It is our task to
show that this space of solutions is isomorphic to the conformal operator space
described in the previous section.

Let us start writing the general solution to the equations
(\ref{fn1})--(\ref{fn3}). It reads
\begin{equation}
F_{n}^{\Phi }(\theta _{1},\ldots ,\theta _{n})=U_{n}^{\Phi }(\theta
_{1},\ldots ,\theta _{n})\prod_{1\leq i<j\leq n}\frac{F_{min}(\theta _{i}-
\theta _{j})}{\cosh \frac{\theta _{i}-\theta _{j}}{2}\left[ \cosh (\theta _{i}-
\theta_{j})+\frac{1}{2}\right] }\text{.}  \label{fn}
\end{equation}
Here the factors in the denominator introduce the bound state and annihilation
poles prescribed by (\ref{fn4}) and (\ref{fn3}), which are the only
singularities of the form factors in rapidity space, while
\begin{equation}
F_{min}(\theta )=-i\sinh \frac{\theta }{2}\,\exp \left\{ 2\int_{0}^{\infty }%
\frac{dt}{t}\frac{\cosh \frac{t}{6}}{\cosh \frac{t}{2}\sinh t}\sin ^{2}\frac{%
(i\pi -\theta )t}{2\pi }\right\}  \label{fmin}
\end{equation}
is the solution of the equations
\begin{equation}
F(\theta )=S(\theta )F(-\theta )
\end{equation}
\begin{equation}
F(\theta +2i\pi )=F(-\theta )
\end{equation}
free of zeros and poles for Im$\theta \in (0,2\pi )$; it behaves asymptotically
as
\begin{equation}
\lim_{|\theta |\rightarrow \infty }e^{-|\theta |}F_{min}(\theta
)=C_{\infty },  \label{asyp-fmin}
\end{equation}
with
\begin{equation}
C_{\infty }=\frac{-1}{4\gamma ^{2}}\,,\hspace{1.5cm}
\gamma =\exp \left\{ 2\int_{0}^{\infty }\frac{dt}{t}\,\frac{\sinh \frac{t}{2}%
\sinh \frac{t}{3}\sinh \frac{t}{6}}{\sinh ^{2}t}\right\} \,.  \label{hn}
\end{equation}

All the information specifying the operator $\Phi $ is contained in
the functions $U_{n}^{\Phi }(\theta _{1},\ldots ,\theta _{n})$. They are
entire functions of the rapidities, symmetric and (up to a factor $%
(-1)^{n-1} $) $2\pi i$-periodic in all $\theta _{j}$'s, and homogeneous of
degree $s_{\Phi }$ (i.e. they account for the property (\ref{fn0})). Of course
the functions $U_n^\Phi$ with different $n$ are related by the residue
equations (\ref{fn4}) and (\ref{fn3}). These equations allow to build a
solution starting from an initial condition for $n=1$, and then
determining the matrix elements with a larger number of particles. In doing
this, however, one should keep in mind that there can be more solutions
corresponding to a given initial condition. Indeed, $N$-particle matrix
elements with vanishing residues on the bound state and kinematical poles are
themselves initial conditions of {\it kernel} solutions which in a linear
combination give no contribution for $n<N$. Enumerating the kernel solutions is
then essential for counting the independent solutions of the form factor
equations, as originally observed in \cite{KM}.

We call \textit{minimal} scalar $N$-kernel solution $K_{n}^{N}(\theta
_{1},...,\theta _{n})$\ the solution of the form factor equations (\ref{fn0}%
)--(\ref{fn3}) with $s_\Phi=0$ and initial condition
\begin{equation}
K_{n}^{N}(\theta _{1},...,\theta _{n})\ =\left\{
\begin{array}{c}
\text{ \ \ \ \ }0\text{ \ \ \ \ \ \ \ \ \ \ \ \ \ \ \ \ \ \ \ \ \ \ \ \ \ \
\ \ for }n<N \\
\\
\prod_{1\leq i<j\leq N}F_{min}(\theta _{i}-\theta _{j})\text{\ \ for }n=N,
\end{array}
\right.  \label{N-kernel}
\end{equation}
where $N\geq 2$. The initial condition of the general spin $s$ $N$-kernel
solution differs from this one by a multiplicative factor which is an
entire function of the rapidities, symmetric and $2\pi i$-periodic in all
$\theta_{j}$'s and homogeneous of degree $s$. After
introducing the elementary symmetric polynomials $\sigma_i^{(n)}$ generated by
\begin{equation}
\prod_{i=1}^{n}(x+x_{i})=\sum_{k=0}^{n}x^{n-k}\sigma _{k}^{(n)}(x_{1},\ldots
,x_{n})\,,
\end{equation}
with $x_{i}\equiv e^{\theta _{i}}$, a basis in the space of $N$-kernel
solutions is provided by the solutions
$K_{n}^{(a_{1},..,a_{N-1}|A)}$ with initial condition
\begin{equation}
K_{n}^{(a_{1},..,a_{N-1}|A)}(\theta _{1},...,\theta _{n})=\left\{
\begin{array}{c}
0\text{ \ \ \ \ \ \ \ \ \ \ \ \ \ \ \ \ \ \ \ \ \ \ \ \ \ \ \ \ \ \ \ \ \
for }n<N \\
\\
(\sigma _{N}^{(N)})^{A}\prod_{1\leq i\leq N-1}(\sigma
_{i}^{(N)})^{a_{i}}K_{N}^{N}(\theta _{1},...,\theta _{N})\text{\ \ for }n=N,
\end{array}
\right.  \label{K-general}
\end{equation}
where $a_1,\ldots a_{N-1}$ are non-negative integers and $A$ is an integer.
If we formally define the spin $s$ '$1$-kernel' solution with initial condition
\begin{equation}
K_{n}^{(s)}(\theta)=\left\{
\begin{array}{c}
0\text{\ \ \ \ \ \ \ \ \ \ \ \ \ \ \ \ for }n=0 \\
\\
\left(\sigma _{1}^{(1)}\right)^s=e^{s\theta}\text{\ \ \ \ \ \ for }n=1,
\end{array}
\right.
\label{1kernel}
\end{equation}
and formally associate the identity solution $F_n^I=\delta_{n,0}$ to $N=0$,
we have that {\it all} the solutions of the form factor equations
(\ref{fn0})--(\ref{fn3}) can be written as linear combinations of the
$N$-kernel solutions with $N\geq 0$. In the following we perform our analysis
of the space of solutions within this basis.

It follows from (\ref{K-general}) and (\ref{1kernel}) that the spin is
\begin{equation}
s=\sum_{i=1}^{N-1}ia_{i}+NA\,.  \label{K_Spin-condition}
\end{equation}
Since (\ref{asyp-fmin}) implies
\begin{equation}
K_{N}^{N}\left( \theta _{1}+\alpha ,..,\theta _{k}+\alpha ,\theta
_{k+1},...,\theta _{N}\right) \sim e^{k(N-k)\alpha }
\label{asymptotic-N-kernel}
\end{equation}
for $\alpha \rightarrow +\infty $, $N>1$ and $1\leq k\leq N-1$, we
have\footnote{In order to simplify the notation the dependence of
$y_k$ on $a_1,\ldots,a_{N-1},A$ is not indicated explicitly.}, in the same
limit and within the same restrictions on $N$ and $k$,
\EQ
K _{N}^{(a_{1},..,a_{N-1}|A)}(\theta_1+\alpha,\ldots,\theta_k+
\alpha,\theta_{k+1},\dots,\theta_{n})\sim e^{y_k\alpha},
\EN
with
\begin{equation}
y_k=\sum_{i=1}^{k-1}ia_{i}+k\left( \sum_{i=k}^{N-1}a_{i}+A+N-k\right)\,.
\label{yk}
\end{equation}
After defining
\EQ
y=\mbox{Max}\{y_{k}\}_{k=\{1,...,N-1\}},
\label{y}
\EN
we can attach to each solution $K_{n}^{(a_{1},..,a_{N-1}|
A)}$ two non-negative integers $l$ and $\bar{l}$ in the following way
\bea
&& l=\mbox{Max}\{s,y,0\},\label{l}\\
&& \bar{l}=l-s\,.\label{lbar}
\eea
By definition, the condition
\EQ
y_k\leq l,
\label{K_Asymptotic-constrains}
\EN
is satisfied and, if both $l$ and $\bar{l}$ are non-vanishing,
there certainly exists at least one value of $k$ for which it holds as an
equality. Since $y_{N-1}=-A+s+N-1$, (\ref{K_Asymptotic-constrains}) with
$k=N-1$ gives $A\geq N-\bar{l}-1$, and then the parameterization
\begin{equation}
A=a_{N}+N-\bar{l}-1,  \label{Kernel-parametrization}
\end{equation}
with $a_{N}$ a non-negative integer.


Then we see that to each $K_{n}^{(a_{1},..,a_{N-1}|A)}$ with $N>1$ in the
basis of kernel solutions we can associate two non-negative intergers, $l$ and
$\bar{l}$, whose difference coincides with the spin. Moreover, taking into
account (\ref{Kernel-parametrization}), each solution is identified by
$N$ non-negative integers $a_1,\ldots,a_N$, and we will use the notation
\EQ
K_{n}^{(a_{1},..,a_{N})}(\theta_1,\ldots,\theta_n)=
K_{n}^{(a_{1},..,a_{N-1}|A)}(\theta_1,\ldots,\theta_n)\,.
\EN

As for the cases $N=0,1$, since the set of $y_k$'s is empty, we set $y\equiv
0$, so that $l$ and $\bar{l}$ are still defined by (\ref{l}) and (\ref{lbar}).
Notice that no $N$-kernel solution with $N>1$ is compatible with $l=\bar{l}=0$.
Indeed, (\ref{K_Spin-condition}) with $s=0$ gives $A\leq 0$, which contradicts
(\ref{Kernel-parametrization}) with $\bar{l}=0$ and $N>1$. Hence, there are
only two independent solutions with $l=\bar{l}=0$\,: the identity, which was
associated to $N=0$, and the solution with $N=1$ and $s=0$.

\resection{Isomorphism between\ critical and off-critical operator space}
We now want to count how many independent solutions of the form factor
equations (\ref{fn0})--(\ref{fn3}) correspond to a given pair of non-negative
integers $l$ and $\bar{l}$ (we say these solutions are of type $(l,\bar{l})$).
In the basis of the $N$-kernel solutions discussed in the previous section the
problem reduces to counting how many sets of integers $a_1,\ldots,a_N$ can
determine the given values of $l$ and $\bar{l}$ through the relations
(\ref{K_Spin-condition}), (\ref{l}) and (\ref{lbar}).

Let us consider first the case of solutions of type $(l,0)$, which is
particularly simple. Since $l=s$, the only constraint comes from
(\ref{K_Spin-condition}) which becomes
\EQ
\sum_{i=0}^{N}ia_{i}=l-N(N-1),
\label{K_right-chiral}
\EN
and holds for all $N\geq 0$. For a given $N$, the number of ways of
satisfying this condition with non-negative integers $a_1,\ldots,a_N$ coincides
with the number of partitions of $l-N(N-1)$ into the positive integers $i=1,
\ldots,N\leq N^{(l)}$, where $N^{(l)}$ is the largest integer ensuring the
non-negativity of $l-N(N-1)$. We saw in section~2 that the number of such
partitions is $P(N,l-N(N-1))$, so that the dimension of the space of solutions
of type $(l,0)$ is
\begin{equation}
d(l,0)=\sum_{N=0}^{\infty }P(N,l-N(N-1))\,.
\end{equation}
Comparison with (\ref{dim-tot(l,0)}) then gives
\begin{equation}
d(l,0)=d_{I}(l)+d_{\varphi }(l),
\end{equation}
i.e. for any non-negative $l$ the dimension of the space of solutions of type
$(l,0)$ of the form factor equations for the massive theory identically
coincides with the total dimension of the space of operators of level $(l,0)$
in the critical theory.

An analogous result holds for the space of solutions of type $(0,
\bar{l})$. Indeed, we show in appendix~A that, for each solution $K_{n}^{(
a_{1},..,a_{N})}(\theta_1,\ldots,\theta_n)$ of type $(l,0)$, a corresponding
solution
$\bar{K}_{n}^{(a_{1},..,a_{N})}(\theta_1,\ldots,\theta_n)$ of type $(0,{l})$
is obtained performing in (\ref{K-general}) the substitution
$\sigma_{i}^{(N)}\to\bar{\sigma}_{i}^{(N)}$, where $\bar{\sigma}_{i}^{(n)}$
stay for the symmetric polynomials computed in $\bar{x}_{i}\equiv
e^{-\theta _{i}}$.

In principle the previous counting procedure can be extended to the solutions
of type $(l,\bar{l})$ with both $l$ and $\bar{l}$ non-vanishing. In this case,
however, the analysis is substantially complicated by the fact that $l$ now
coincides with $y$, which is non-trivially determined through (\ref{yk}) and
(\ref{y}). There is, however, a simpler path. We show in appendix~B that
a solution $K_{n}^{(a_{1},..,a_{N})}$ of type $(l,\bar{l})$ satisfies the
asymptotic factorization property
\begin{eqnarray}
&& \lim_{\alpha \rightarrow +\infty }e^{-l\alpha
}K_{N}^{(a_{1},..,a_{N})}(\theta _{1}+\alpha ,...,\theta _{R}+\alpha
,\theta _{R+1}...,\theta _{N})=\notag \\
&& \hspace{3cm}(C_{\infty
})^{RL}K_{R}^{(a_{1}^{(l)},..,a_{R}^{(l)})}(\theta _{1},...,\theta
_{R})\,\bar{K}_{L}^{(a_{1}^{(\bar{l})},..,a_{L}^{(\bar{l})})}(\theta
_{R+1}...,\theta _{N})\,,  \label{Kernel-factorization-(l,lbar)}
\end{eqnarray}
where $K_{R}^{(a_{1}^{(l)},..,a_{R}^{(l)})}$ defines a solution of type
$(l,0)$, $\bar{K}_{L}^{(a_{1}^{(\bar{l})},..,a_{L}^{(\bar{l})})}$ defines
a solution of type $(0,\bar{l})$, $1\leq R\leq N-1$, $L=N-R$, and the integers
$a_{1}^{(l)},..,a_{R}^{(l)}$, $a_{1}^{(\bar{l})},..,a_{L}^{(\bar{l})}$
are determined by the $a_{1},..,a_{N}$ as
\begin{eqnarray}
a_{i}^{(l)} &=&a_{i}\text{ \ \ \ for }\,\,1\leq i\leq R-1,
\label{Coeff-R_non-chiral} \\
&&  \notag \\
a_{N-i}^{(\bar{l})} &=&a_{i}\text{ \ \ \ for }\,\,R+1\leq i\leq N-1,
\label{Coeff-L_non-chiral}
\end{eqnarray}
\begin{equation}
a_{R}^{(l)}=\sum_{i=R}^{N}a_{i}-\bar{l}+2L\,,\hspace{1cm}a_{L}^{(\bar{l%
})}=-\sum_{i=R+1}^{N}a_{i}+\bar{l}-2(L-1).  \label{a_r^l,a_L^lbar}
\end{equation}
Inversely, the specification of the $a_i$'s in terms of the $a_i^{(l)}$'s and
$a_i^{(\bar{l})}$'s is completed by the relations
\begin{eqnarray}
a_{N} &=&-\sum_{i=1}^{L}a_{i}^{(\bar{l})}+\bar{l}-2(L-1), \label{a_n}\\
&&  \notag \\
a_{R} &=&a_{R}^{(l)}+a_{L}^{(\bar{l})}-2. \label{a_r}
\end{eqnarray}
The non-negativity of $a_R$ implies
\begin{equation}
a_{R}^{(l)}+a_{L}^{(\bar{l})}\geq 2\,,  \label{cond-R_L_composition}
\end{equation}
while that of $a_N$ follows from
\begin{equation}
a_{N}\geq -\sum_{i=1}^{L}ia_{i}^{(\bar{l})}+\bar{l}-2(L-1)=(L-2)(L-1)\geq 0\,.
\end{equation}

Hence, equation (\ref{Kernel-factorization-(l,lbar)}) can be used to
characterize all solutions of type $(l,\bar{l})$ in terms of those of type
$(l,0)$ and $(0,\bar{l})$. More precisely, given a solution of type
$(l,0)$ and one of type $(0,\bar{l})$, they specify a solution
of type $(l,\bar{l})$ provided (\ref{cond-R_L_composition}) is satisfied.
The two conditions
\bea
&& a_{R}^{(l)}=0,\hspace{1cm} a_{L}^{(\bar{l})}\geq 2,\\
&& a_{R}^{(l)}\geq 1,\hspace{1cm}a_{L}^{(\bar{l})}\geq 1,
\eea
exhaust all the independent possibilities and lead to the following expression
for the dimension of the space of solutions of type $(l,\bar{l})$
\begin{equation}
d(l,\bar{l})=d(0|(l,0))d(2|(0,\bar{l}))+d(1|(l,0))d(1|(0,\bar{l})),
\label{dim-(l,lbar)}
\end{equation}
where $d(0|(l,0))$ is the dimension of the subspace of solutions spanned by
the $R$-kernels of type $(l,0)$ with $a_{R}^{(l)}=0 $ and $R\leq N^{(l)}$, and
$d(i|(l,0))$ with $i\geq 0$ is the dimension of the subspace of solutions
spanned by the $R$-kernels of type $(l,0)$ with $a_{R}^{(l)}\geq i$ and
$R\leq N^{(l)}$ (analogous definitions for the dimensions of subspaces of type
$(0,\bar{l})$ are understood). The formula for $d(i|(l,0))$ with $i\geq 0$ is
simply derived redefining $a_{R}^{(l)}=a_{R}^{(l),i}+i$, with now
$a_{R}^{(l),i}\geq 0$, so that
\begin{equation}
d(i|(l,0))=\sum_{N=0}^{\infty }P(N,l-N(N+i-1)).
\end{equation}
By definition $d(0|(l,0))$ gives the number of partitions of $l-R(R-1)$ into
the integers $1,\ldots,R-1$ $(a_{R}^{(l)}=0)$, so that
\begin{equation}
d(0|(l,0))=\sum_{N=1}^{\infty }P(N-1,l-N(N-1)),
\end{equation}
which implies $d(0|(l,0))=d(2|(l,0))$. Finally, recalling (\ref{d-i-phi})
we obtain the identities
\begin{equation}
d(2|(l,0))=d_{I}(l),\text{ \ \ }d(1|(l,0))=d_{\varphi }(l),
\end{equation}
so that (\ref{dim-(l,lbar)}) becomes
\begin{equation}
d(l,\bar{l})=d_{I}(l)d_{I}(\bar{l})+d_{\varphi }(l)d_{\varphi }(\bar{l})\,.
\label{dllbar}
\end{equation}
This formula completes our proof showing that the space of solutions of the
form factor equations for the massive Lee-Yang model decomposes into
subspaces labeled by pairs of non-negative integers $l$ and $\bar{l}$ whose
dimensionality coincides with that of the subspace of conformal operators of
level $(l,\bar{l})$.

\resection{Conclusion}
We have shown in this paper for the Lee-Yang model how the operator space
reconstructed from the particle dynamics of the massive theory through the
form factor equations (\ref{fn0})--(\ref{fn3}) is a direct
sum of subspaces with given levels which exactly coincides with the
decomposition dictated by conformal symmetry at the fixed point.

It is worth stressing that we are able to achieve this result because, through
(\ref{l}) and (\ref{lbar}), we are able to attach to each solution belonging to
a basis for the whole space of solutions of the form factor equations two
non-negative integers that, after the isomorphism has been shown, is natural
to call levels. This notion of levels for the generic form factor solution is
absent in previous investigations, and this is why an equation like
(\ref{dllbar}) is not contained there.

At criticality conformal symmetry also naturally yields the notion of
operator families corresponding the lowest weight representations of the
Virasoro algebra. In the massive theory, in absence of internal symmetries
which distinguish them, the two operator families of the Lee-Yang model cannot
be disentangled using the form factor equations (\ref{fn0})--(\ref{fn3}) only.
Additional information is needed to pass from the classification of the
solutions according to the levels to the identification of specific operators
within them.

As far as the family of $\varphi$ is concerned, this operator, being
responsible for the breaking of conformal symmetry, is proportional to the
trace $\Theta$ of the energy-momentum tensor, and on this ground the solution
of the form factor equations corresponding to it was originally identified in
\cite{Smirnovreduction,Alyosha}. This solution coincides with that
with $N=1$, $s=0$ in the $N$-kernel basis that in section~3 we identified as
the only operator other than the identity with $l=\bar{l}=0$.

The first non-trivial representatives of the identity family appear at level
$2$. The solutions for the energy-momentum components $T$ and $\bar{T}$ with
levels $(2,0)$ and $(0,2)$, respectively, are immediately obtained from the
solution for $\Theta$ through the energy-momentum conservation equations.
Hence, the first genuinely new solution in this family is that for the
composite operator $T\bar{T}$ with levels $(2,2)$. Here it is worth recalling
that, while at the conformal point the decoupling of holomorphic and
anti-holomorphic components reduces non-chiral operators to trivial products of
the chiral ones, in the massive theory the decoupling is lost\footnote{Equation
(\ref{Kernel-factorization-(l,lbar)}) expresses that the decoupling is
recovered in the conformal, high energy limit.} and non-chiral
operators need to be suitably defined as regularized products. It is then
particularly relevant that our proof of one-to-one correspondence between
operators at and away from criticality includes the non-chiral ones.
The form factor solution for $T\bar{T}$ in the Lee-Yang model was determined in
\cite{ttbar} exploiting also some general properties of this operator obtained
in \cite{Sasha}.

In \cite{seven} all the operators with $l,\bar{l}\leq 7$ for the massive
Lee-Yang model were obtained acting on the form factor solutions for $\Theta$,
$T$, $\bar{T}$ and $T\bar{T}$ with the first few conserved quantities of this
integrable quantum field theory. For these values of $l$ and $\bar{l}$,
(\ref{dllbar}) was then reproduced with the two terms in the r.h.s.
disentangled.

It is reasonable to expect that the approach illustrated here for the
simplest non-trivial case can be generalized to more complicated integrable
quantum field theories. In all massive integrable cases the analysis of the
structure of the operator space reduces to the study of the space of solutions
of a system of equations like (\ref{fn0})--(\ref{fn3}), complicated in general
by the presence of several species of particles. The form factor equations,
instead, undergo substantial modifications when integrability is lost and
particle production becomes possible. In the general two-dimensional case,
however, both integrable and non-integrable renormalization group trajectories
originate from a given fixed point. Then, up to symmetry breaking effects,
the non-integrable form factor equations should yield the same operator
space than the integrable ones.

\vspace{1cm}\textbf{Acknowledgments.} The work of G.D. is partially supported
by the ESF grant INSTANS and by the MUR project ``Quantum field theory and
statistical mechanics in low dimensions''. The work of G.N. was supported by
the ANR program MIB-05 JC05-52749 and is currently supported by the contract
MEXT-CT/2006/042695.

\appendix

\resection{Chiral and antichiral solutions}
Notice that the generic $N$-kernel $K_{n}^{(a_{1},..,a_{N})}(\theta
_{1},...,\theta _{n})$ of type $(0,\bar{l}>0)$ can be rewritten as
\begin{equation}
\bar{K}_{n}^{(\bar{a}_{1},..,\bar{a}_{N})}(\theta _{1},...,\theta
_{n})=\left\{
\begin{array}{c}
0\text{ \ \ \ \ \ \ \ \ \ \ \ \ \ \ \ \ \ \ \ \ \ \ \ \ \ \ \ \ \ \ \ \ \
for }n<N, \\
\\
(\bar{\sigma}_{N}^{(N)})^{N-1}\prod_{1\leq i\leq
N}(\bar{\sigma}_{i}^{(N)})^{%
\bar{a}_{i}}K_{N}^{N}(\theta _{1},...,\theta _{N})\text{\ \ for }n=N,
\end{array}
\right.   \label{K-bar}
\end{equation}
where $\bar{\sigma}_{i}^{(N)}$ stay for the symmetric polynomials computed in
$\bar{x}_{i}\equiv e^{-\theta _{i}}$, and that the integers $\bar{a}_{1},..,
\bar{a}_{N}$ given by
\EQ
\bar{a}_{i}=a_{N-i},\hspace{.5cm}n<N;\hspace{1cm}
\bar{a}_{N}=-\left(\sum_{i=1}^{N}a_{i}+2(N-1)-\bar{l}\right)   \label{a-bar}
\end{equation}
are non-negative. This is obvious for
$\bar{a}_{i}$ with $n<N$, while for $\bar{a}_{N}=-y_{1}(a_{1},...,a_{N})$ it
follows from (\ref{K_Asymptotic-constrains}) with $l=0$. In addition, in terms
of the $\bar{a}_{i}$ the condition $s=-\bar{l}$ is rewritten as
\begin{equation}
\sum_{i=1}^{N}i\bar{a}_{i}=\bar{l}-N(N-1).  \label{K_left-chiral}
\end{equation}

Inversely, any solution $\bar{K}_{n}^{(\bar{a}_{1},..,\bar{a}_{N})}(\theta
_{1},...,\theta _{n})$ with $\bar{a}_{1},..,\bar{a}_{N}$ non-negative
integers satisfying $(\ref{K_left-chiral})$ is a $N$-kernel of type
$(0,\bar{l})$. Indeed, due to $(\ref{K_left-chiral})$,
$\bar{K}_{n}^{(\bar{a}_{1},..,\bar{a}_{N})}(\theta _{1},...,\theta_{n})$
has spin $s=-\bar{l}$, while using $(\ref{a-bar})$ $y_{k}$ can be rewritten
in terms of $\bar{a}_{1},..,\bar{a}_{N}$ as
\begin{equation*}
y_{k}=-\left( \sum_{i=0}^{k-1}(k-i)\bar{a}_{N-i}+k(k-1)\right),
\end{equation*}
so that $y_{k}\leq 0$ for $1\leq k\leq N-1$; then (\ref{l}) implies $l=0$.

Hence we see that the spaces of kernel solutions of type $(\bar{l},0)$ and
$(0,\bar{l})$ are isomorphic. Indeed, $N$ non-negative integers
$\bar{a}_{1},..,\bar{a}_{N}$ satisfying $(\ref{K_left-chiral})$
define the solution $K_{n}^{(\bar{a}_{1},..,\bar{a}_{N})}(\theta _{1},...,
\theta _{n})$ of type $(\bar{l},0)$ (as discussed in section~4) as well as the
solution $\bar{K}_{n}^{(\bar{a}_{1},..,\bar{a}_{N})}(\theta_{1},...,
\theta _{n})$ of type $(0,\bar{l})$.

\resection{Asymptotic factorization}

It follows from (\ref{asyp-fmin}) that the minimal $N$-particle kernel
(\ref{N-kernel}) satisfies the asymptotic factorization property
\begin{equation}
\lim_{\alpha \rightarrow +\infty }e^{-RL\alpha }K_{N}^{N}(\theta _{1}+\alpha
,...,\theta _{R}+\alpha ,\theta _{R+1}...,\theta _{N})=(C_{\infty
})^{RL}K_{R}^{R}(\theta _{1},...,\theta _{R})K_{L}^{L}(\theta
_{R+1}...,\theta _{N}).
\end{equation}
On the other hand the elementary symmetric polynomials enjoy the properties
\begin{align}
\lim_{\alpha \rightarrow +\infty }e^{-k\alpha }\sigma _{p}^{(N)}\left(
x_{1}e^{\alpha },..,x_{k}e^{\alpha },x_{k+1},..,x_{N}\right) & =\sigma
_{k}^{(k)}(x_{1},\ldots ,x_{k})\sigma _{p-k}^{(N-k)}(x_{k+1},\ldots ,x_{N}),%
\hspace{0.1in}k\leq p\leq N,  \label{Sigma_Asympt-Factorization-1} \\
&  \notag \\
\lim_{\alpha \rightarrow +\infty }e^{-p\alpha }\sigma _{p}^{(N)}\left(
x_{1}e^{\alpha },..,x_{k}e^{\alpha },x_{k+1},..,x_{N}\right) & =\sigma
_{p}^{(k)}(x_{1},\ldots ,x_{k}),\hspace{0.1in}\hspace{0.1in}\hspace{0.1in}%
p\leq k\leq N.  \label{Sigma_Asympt-Factorization-2}
\end{align}
Using these equations it is simple to see that
(\ref{Kernel-factorization-(l,lbar)}) holds for a $N$-kernel solution
satisfying (\ref{K_Asymptotic-constrains}) as an equality for $k=R$. However,
we still have to prove that the factors on the r.h.s. of (\ref
{Kernel-factorization-(l,lbar)}) are indeed a $R$-kernel of type $(l,0)$ and
a $L$-kernel of type $(0,\bar{l})$, i.e. that the integers
$a_{1}^{(l)},..,a_{R}^{(l)}$
and $a_{1}^{(\bar{l})},..,a_{L}^{(\bar{l})}$ are non-negative and satisfy
the condition (\ref{K_right-chiral}) and (\ref{K_left-chiral}), respectively.
Equation (\ref{K_right-chiral}) for the $a_{i}^{(l)}$ follows from
\begin{equation}
\sum_{i=1}^{R}ia_{i}^{(l)}=\sum_{i=1}^{R-1}ia_{i}+R(\sum_{i=R}^{N}a_{i}-
\bar{l}+2L)=l-R(R-1),
\label{a4}
\end{equation}
where the last equality is due to the fact that (\ref
{K_Asymptotic-constrains}) holds as an equality for $k=R$.
Analogously,
\begin{equation}
\sum_{i=1}^{L}ia_{i}^{(\bar{l})}=-\sum_{j=R+1}^{N}(j-R)a_{j}+L(\bar{l}%
-2(L-1))=\bar{l}-L(L-1),
\label{a5}
\end{equation}
where we have used
\begin{equation}
\sum_{j=R+1}^{N}(j-R)a_{j}=-\bar{l}+L(\bar{l}+1-L),
\end{equation}
a result which follows taking the difference of (\ref{K_Spin-condition}) and
(\ref{K_Asymptotic-constrains}) for $k=R$.

The integers $a_{i}^{(l)}$ and $a_{j}^{(\bar{l})}$ with $1\leq i\leq R-1$
and $1\leq j\leq L-1$ are non-negative by (\ref{Coeff-R_non-chiral})-(\ref
{Coeff-L_non-chiral}) and so we have to prove only the non-negativity of $%
a_{R}^{(l)}$ and $a_{L}^{(\bar{l})}$. If we rewrite
\begin{equation*}
a_{R}^{(l)}=(\sum_{i=1}^{R-1}ia_{i}+R\sum_{i=R}^{N}a_{i})-(%
\sum_{i=1}^{R-1}ia_{i}+(R-1)\sum_{i=R}^{N}a_{i})-\bar{l}+2L,
\end{equation*}
we get $a_{R}^{(l)}\geq 0$ using (\ref{K_Asymptotic-constrains}) for $k=R$ and
$k=R-1$ . Similarly,
\begin{equation*}
a_{L}^{(\bar{l})}=(\sum_{i=1}^{R}ia_{i}+R\sum_{i=R+1}^{N}a_{i})-(%
\sum_{i=1}^{R}ia_{i}+(R+1)\sum_{i=R+1}^{N}a_{i})+\bar{l}-2(L-1),
\end{equation*}
so using (\ref{K_Asymptotic-constrains}) for $k=R$ and $k=R+1$ we get
$a_{L}^{(\bar{l})}\geq 0$, in this way completing the proof.

Let us now prove the characterization of kernel solutions of type $(l,\bar{l})$
in terms of those of type $(l,0)$ and $(0,\bar{l})$. We have just to prove
that, given $R$ non-negative integers $a_{1}^{(l)},..,a_{R}^{(l)}$ and $L$
non-negative integers $a_{1}^{(\bar{l})},..,a_{L}^{(\bar{l})}$ satisfying
(\ref{cond-R_L_composition}) and,
respectively, the conditions (\ref{K_right-chiral}) and (\ref{K_left-chiral}),
then the integers $a_{1},..,a_{N}$ determined by
(\ref{Coeff-R_non-chiral}), (\ref{Coeff-L_non-chiral}),
(\ref{a_n}), (\ref{a_r})
satisfy (\ref{K_Spin-condition}) and (\ref{K_Asymptotic-constrains}), with
(\ref{K_Asymptotic-constrains}) which is an equality for $k=R$. For this
purpose notice that the difference of (\ref{a4}) and (\ref{a5}) gives
\begin{equation}
\sum_{i=1}^{N}ia_{i}=\sum_{i=1}^{R}ia_{i}^{(l)}-\sum_{j=1}^{L}ja_{j}^{(\bar{l%
})}-N(2L-(\bar{l}+1))+R-L=l-\bar{l}-N(N-(\bar{l}+1)),
\end{equation}
where to derive the last equality we have used (\ref{K_right-chiral}) for
the $a_{i}^{(l)}$ and (\ref{K_left-chiral}) for the
$a_{j}^{(\bar{l})}$. This is the spin condition (\ref{K_Spin-condition}).
The identity (\ref{K_Asymptotic-constrains}) for $k=R$ follows from
\begin{equation}
\sum_{i=1}^{R}ia_{i}+R\sum_{i=R+1}^{N}a_{i}=\sum_{i=1}^{R}ia_{i}^{(l)}-R(2L-%
\bar{l})=l-R(N-R+(N-(\bar{l}+1))),
\end{equation}
where to derive the last equality we have used the condition (\ref
{K_right-chiral}) for the $a_{i}^{(l)}$.

Consider now the inequality (\ref{K_Asymptotic-constrains}) for $k\neq R$. For
$k<R$ we have
\begin{equation}
\sum_{i=1}^{k-1}ia_{i}+k\sum_{i=k}^{N}a_{i}=\sum_{i=1}^{k-1}ia_{i}^{(l)}+k%
\sum_{i=k}^{R}a_{i}^{(l)}-k(2L-\bar{l})\leq l-k((N-k)+(N-(\bar{l}+1))),
\label{K_Asymptotic-constrains-k<r}
\end{equation}
where to derive the last inequality we have used the condition (\ref
{K_Asymptotic-constrains}) with $k<R$ for the integers
$a_{1}^{(l)},..,a_{R}^{(l)}$. For $k>R$ we have
\begin{equation}
\sum_{i=1}^{k-1}ia_{i}+k\sum_{i=k}^{N}a_{i}=\sum_{i=1}^{R}ia_{i}^{(l)}-%
\sum_{j=1+(N-k)}^{L}(j-N+k)a_{j}^{(\bar{l})}-2R-k(2(L-1)-\bar{l})\leq
l-k((N-k)+(N-\bar{l}-1)),  \label{K_Asymptotic-constrains-k>r}
\end{equation}
where to obtain the last inequality we have used (\ref{K_right-chiral}) for
the $a_{i}^{(l)}$ and
\begin{equation}
\sum_{j=1+p}^{L}(j-p)a_{j}^{(\bar{l})}\geq -(L-p)(p+(L-1))\,;
\end{equation}
this last inequality follows taking the difference of (\ref{K_left-chiral})
and (\ref{K_Asymptotic-constrains}) applied to the $a_{i}^{(\bar{l})}$.
(\ref{K_Asymptotic-constrains-k<r}) and (\ref{K_Asymptotic-constrains-k>r})
provide the conditions (\ref{K_Asymptotic-constrains}) for $k\neq R$, in this
way completing the proof.


\end{document}